\documentclass[pre,twocolumn,amsmath,showpacs,floatfix]{revtex4}
\usepackage{amssymb}
\usepackage{bm}
\usepackage{graphicx}
\usepackage{array}
\usepackage{hhline}
\usepackage{longtable}
\usepackage{comment}

\begin{document}

\title{Theory of depletion--induced phase transition from chiral smectic--$A$ twisted ribbons to semi--infinite flat membranes}
\author{C. Nadir Kaplan,$^{1}$ Hao Tu,$^{2}$ Robert A. Pelcovits,$^{2}$ and Robert B. Meyer$^{1}$}
\affiliation{$^1$The Martin Fisher School
of Physics, Brandeis University, Waltham, Massachusetts 02454} 
\affiliation{$^2$Department of Physics, Brown University, Providence, Rhode Island 02912}

\date{\today}

\begin{abstract}
We consider a theoretical model for the chiral smectic $A$ twisted ribbons observed in assemblies of $fd$ viruses condensed by depletion forces. The depletion interaction is modeled by an edge energy assumed to be proportional to the depletant polymer in solution. Our model is based on the Helfrich energy for surface bending and the de Gennes model of chiral smectic $A$ liquid crystals with twist penetration at the edge.  We consider two variants of this model, one with the conventional Helfrich Gaussian curvature term, and a second with saddle--splay energy. A mean field analysis of both models yields a first--order phase transition between ribbons and semi--infinite flat membranes as the edge energy is varied. The phase transition line and tilt angle profile are found to be nearly identical for the two models; the pitch of the ribbon, however, does show some differences. Our model yields good qualitative agreement with experimental observations if the sign of the Gaussian curvature or saddle--splay modulus is chosen to favor negative Gaussian curvature. 
\end{abstract}

\pacs{61.30.Dk,64.70.M-,02.40.Hw}

\maketitle

\def\s{\rule{0in}{0.28in}}

\section{Introduction}

Smectic $A$ (Sm-$A$) layers expel twist and bend deformations,
just as magnetic fields are expelled from bulk superconductors
\cite{DeGennes1,DeGennes2}. In analogy with the London penetration
length, the typical distance to which a magnetic field penetrates
into a superconductor, two penetration depths can be defined in
Sm-$A$ liquid crystals describing respectively the penetration of
twist and bend deformations at the edges or around isolated
defects of Sm-$A$ layers. When a Sm-$A$ sample is composed of
chiral molecules (molecules without mirror symmetry), twist
deformations appear intrinsically, driving the formation of novel
structures, due to the competition of the twist deformations with
the tendency of the molecules to build a perfect twist--free
smectic layer~\cite{Selinger}.  When the layer forming tendency dominates, the
twist is restricted to a band at the layer edge, but when the
twist deformations are strong enough, simple flat membranes are
replaced by a variety of twisted structures, including twisted
ribbons, double helices, arrays of twist walls in membranes, and
periodic arrays of pores.  In this paper, we study a theoretical
model for one result of this basic competition of ordering forces,
the transition between flat disks and twisted ribbons.

Recently, there have been both experimental and theoretical
studies to visualize and quantify the effects of twist and bend
deformations on chiral smectic $A$ (Sm-$A^\ast$) single-layer
membranes in the weak chirality limit, as well as investigating
the relative stability of different geometries of membranes seen
in experiments. These membranes are formed under certain
conditions in aqueous solutions of filamentous virus particles and
a non-binding polymer, which acts to condense the virus particles
into dense phases by depletion forces.  Barry \textit{et. al.}
have studied flat Sm-$A^\ast$ monolayer membranes in the form of
disks composed of rodlike \textit{fd} virus particles to measure
the twist penetration length $\lambda_t$ at the edges of the disks
\cite{Barry}. The persistence length of these viruses is
$2.8\pm0.7~\mu m$ and the length of $fd$ viruses was around 1 $\mu
m$, making them nearly rigid rods \cite{Dogic3}. Because they act
like hard rods, the condensed phases they form due to the
depletion force are mainly entropy driven \cite{Dogic1,Dogic2}.
These viruses form cholesteric and Sm-$A^\ast$ phases with
increasing concentration. The radii of the disks studied were tens
of micrometers. Due to the length scale defined by the particle
length, the twist penetration could be visualized using optical
microscopy. Due to the tilting of the virus particles relative to
the layer normal, in the twist penetration region, the twist could
be measured by the resulting change in retardance of transmitted
polarized light. The twist penetration length $\lambda_t$ was determined to be about 0.5 $\mu
m$. Because the disks were relatively large compared to this, the
twist penetration profile fit well to the analytic theory of twist
penetration for a semi--infinite membrane, with the director tilting tangential to the edge of the membrane. From the estimated
maximum tilt at the disk edge, the product of penetration length
and cholesteric wave vector $q$ was estimated as approximately
$q\lambda_t=0.71$, for the particular sample studied.

In analyzing the energy of the edge of a disk, there are two
important terms, the bare edge energy, or line tension, of the
edge, $\gamma$, and the net reduction of free energy in the
material near the edge due to the twist penetration.  The line
tension is controlled by the concentration of depletant polymer in
solution.  As this concentration is lowered, $\gamma$ is reduced.
In a simple analysis, if the second term then becomes larger than
the first, then the net energy associated with the edge is
negative, and the material will adopt a structure that maximizes
the length of edge, relative to the area of flat disk. This can be
achieved in several ways.

One mechanism, previously studied, is a transition from semi--infinite
membranes to small disks as the structure that minimizes the mean free
energy density of the system.  The mean free energy density of
small Sm-$A^\ast$ disks, which have both twist and bend
deformations at the curved disk edge, has been calculated
\cite{Pelcovits}. It is found that small disks with a twist/bend
penetration length on the order of the disk radius are at least
metastable relative to large disks upon reducing the line tension
of the edge. The magnitude of the critical line tension at which a
second--order phase transition occurs from semi--infinite membranes to small
disks is affected by the magnitudes of the twist wave vector $q$
and the twist penetration length $\lambda_t$.

Rather than simple small disks, a diversity of structures is
observed experimentally to replace large disks as the
concentration of depletion agents is lowered \cite{Gibaud}.
Twisted ribbons (minimal surfaces to a double helix) are commonly
seen.  In this paper we carry out a mean--field study of the
twisted ribbon structure. Throughout this work we assume that the
smectic order parameter coherence length $\xi$ is less than
$\lambda_t$, so that the Sm-$A^\ast$ phase is analogous to a type
II superconductor. Additionally, we focus here on a model in which
twist and bend deformations near the edges of the Sm-$A^\ast$
layers are assumed to be driven by the chirality of the molecules,
as a continuation of the previous work \cite{Barry, Pelcovits}.
%This assumption is justified by the electron microscopy images for effectively infinite disks \cite{Gibaud}.
We obtain the twist penetration profile as a function of the width
of the twisted ribbon. Furthermore, we are able to study the
stability of these structures and determine the first--order phase
transition from isolated Sm-$A^\ast$ twisted ribbons to semi--infinite
flat layers.

The present study is organized as follows: In the next section we
introduce our model for the elasticity of smectic $A^\ast$
membranes and then apply it to twisted ribbons.  In Sec. III, we
use our model to study the transition between infinite disks and
twisted ribbons. We offer concluding remarks in the final section
of the paper.

\section{Free Energy of S\lowercase{m}-$A^\ast$ Membranes}

\subsection{Elasticity of membranes}

We model Sm-$A^\ast$ membranes using the Helfrich model~\cite{Helfrich1,Helfrich2} for the surface bending energy and the de Gennes model~\cite{DeGennes1,DeGennes2} for the Sm-$A$ phase generalized to include chirality. Our model is an extension of one introduced in Ref.~\cite{Seifert} to include variations in the tilt angel of the director relative to the normal to the surface of the membrane. We write the free energy $F$ of an Sm-$A^\ast$ membrane as follows: 
\begin{equation}
\label{eq:7}     %{eq:12b}
F=\int (f_H+f_n) dA+\gamma\oint dl\,,
\end{equation}
where $f_H$ and $f_n$ are the Helfrich and de Gennes free energy densities respectively, $dA$ is the area element of the membrane, and  $dl$ is the arc element length of the edge. The last term represents an edge energy with energy per unit length (line tension) $\gamma$. The Helfrich energy density $f_H$ is given by:
\begin{equation}
\label{eq:2}
f_H=\frac{1}{2}k\left(2H\right)^2+\bar{k}K_G,
\end{equation}
where $H$ and $K_G$ are the mean and the Gaussian curvatures of the surface respectively, $k$ is the bending rigidity and $\bar{k}$ is the Gaussian curvature modulus.  There is no spontaneous curvature in Sm-$A^\ast$ membranes given their up--down symmetry, and thus we have not included such a term. Furthermore, twisted ribbons have zero mean curvature so henceforth we set $H=0$. In terms of $\hat{\mathbf N}$ the local unit normal vector to the surface the Gaussian curvature is given by \cite{Kamien}:
\begin{equation}
K_G=\frac{1}{2} \nabla\cdot\left(\hat{\mathbf N}(\nabla\cdot\hat{\mathbf N})-(\hat{\mathbf N}\cdot\nabla)\hat{\mathbf N}\right)\label{KG}
\end{equation}
If the molecular director field ${\mathbf n}$ is everywhere parallel to the layer normal $\hat{\mathbf N}$, then the Gaussian curvature is equal to the Frank saddle--splay energy density \cite{Kleman,Kamien}. In Sec.~\ref{ss} we consider an alternate model of the membrane where we replace the Gaussian curvature by saddle splay; we compare the results of the two models of a twisted ribbon in Sec.~\ref{results}.

The de Gennes free energy density for the nematic--Sm-$A^\ast$  phase transition is given by:
\begin{equation}\label{NA}
\begin{split}
f_{NA}= &\frac{r}{2} \vert\Psi\vert^2+\frac{u}{4}\vert\Psi\vert^4+D\vert(\nabla-\frac{2\pi i}{d }\delta \mathbf n)\Psi\vert^2\\&+\frac{1}{2}K_1\left(\nabla\cdot\mathbf n\right)^2+\frac{1}{2}K_2 (\mathbf n\cdot\nabla\times\mathbf n-q)^2\\&+\frac{1}{2}K_3 (\mathbf n\times\nabla\times\mathbf n)^2
\end{split}
\end{equation}
Here $\Psi$ is the smectic order parameter, $K_1,K_2$ and $K_3$ denote the splay, twist, and bend Frank elastic constants respectively, $q$ is magnitude of the spontaneous twist wave vector arising from molecular chirality, and  $\delta \mathbf n$ is the deviation of the director from the layer normal. In a bulk Sm-$A$ phase twist and bend distortions which are incompatible with constant interlayer spacing will be expelled; however, as noted by de Gennes both twist and bend can be introduced  at layer edges or around defects by allowing local tilting of $\mathbf n$ relative to the layer normal. Two penetration lengths can be introduced, $\lambda_2=(K_2/D)^{1/2}$ and $\lambda_3=(K_3/D)^{1/2}$, describing respectively the penetration lengths of twist and bend into the smectic phase.

In the limit where the smectic order parameter $\Psi$ is uniform and nonzero (i.e., deep in the smectic $A^\ast$ phase) the de Gennes free energy density reduces (up to a constant) to: 
\begin{equation}
\label{eq:1}
\begin{split}
f_n&= \frac{1}{2}K_1\left(\nabla\cdot\mathbf n\right)^2+\frac{1}{2}K_2 (\mathbf n\cdot\nabla\times\mathbf n-q)^2\\&+\frac{1}{2}K_3 (\mathbf n\times\nabla\times\mathbf n)^2+\frac{1}{2}C\sin^2{\theta}\,,
\end{split}
\end{equation}
where $\theta$ is the tilt angle of the director with respect to the layer normal and the tilt energy density $C$ is proportional to the coupling $D$ appearing in Eq.~\eqref{NA}. 

In the single Frank elastic constant approximation, $K\equiv K_1=K_2=K_3$, which we will henceforth employ in the analysis of Eq.~\eqref{eq:7}, $f_n$ is given by:
\begin{equation}
\label{eq:8}    %{eq:13}
\begin{split}
f_n=\frac{1}{2} K&\left[ \left(\nabla\cdot \mathbf n\right)^2-2q\mathbf n\cdot\left(\nabla\times \mathbf n\right)+\left(\nabla\times\mathbf n\right)^2+q^2\right]\\&+\frac{1}{2} C \sin^2{\theta}.
\end{split}
\end{equation}

Mathematically, a membrane can be represented as a surface with position vector $\mathbf{Y}=\mathbf{Y}(u_1, u_2)$ embedded in a three--dimensional Euclidean space, where $u_1$ and $u_2$ are real variables parameterizing the surface. To calculate the free energy of the membrane we need the following geometric quantities \cite{ZhongCan2,ZhongCan1}: 
\begin{equation}
\label{eq:4}
\begin{split}
&\mathbf{Y}_i\equiv\partial_i \mathbf{Y},\quad \mathbf{Y}_{ij}\equiv\partial_i\partial_j\mathbf{Y}=\Gamma^k_{ij}\mathbf{Y}_k+L_{ij}\hat{\mathbf N},\\
&\partial_i\hat{\mathbf N}=\partial_{N} \mathbf{Y}_i \equiv-L_{ij} g^{jk} \mathbf{Y}_k,\\
&g_{ij}\equiv\mathbf{Y}_i\cdot\mathbf{Y}_j\quad g^{ij}\equiv(g_{ij})^{-1}, \quad g\equiv\det{g_{ij}},\\
&L_{ij}\equiv \mathbf{Y}_{ij}\cdot\hat{\mathbf N},\quad L^{ij}\equiv(L_{ij})^{-1}, \quad L\equiv\det{L_{ij}}.
\end{split}
\end{equation}
The indices $i,j,k=1,2$ (repeated indices being summed over), $\partial_i\equiv \partial_{u_i}$, and $\partial_{N}$ denotes the partial derivative in  the normal direction $\hat{\mathbf N}$.  The completely symmetric tensors $g_{ij}$ and $L_{ij}$ are the first and second fundamental forms of the surface, respectively. The Christoffel symbols $\Gamma^k_{ij}$ are defined by the relation $\Gamma^k_{ij}=g^{km} \mathbf{Y}_{ij}\cdot\mathbf{Y}_m$, and obey the symmetry property $\Gamma^k_{ij}=\Gamma^k_{ji}$.  The unit normal vector of the surface is given by
\begin{equation}
\label{eq:5}    %{eq:12} 
\hat{\mathbf N}=\frac{\mathbf{Y}_1\times\mathbf{Y}_2}{\sqrt{g}}.
\end{equation}
The Gaussian curvature $K_G$, is given by:
\begin{equation}
\label{eq:6} 
K_G=\frac{L}{g},
\end{equation}
and the surface area element is given by $dA=\sqrt{g} du_1 du_2$.

The director field $\mathbf n$ can be expressed in the local basis $\lbrace \mathbf{Y}_1,\mathbf{Y}_2,\hat{\mathbf N}\rbrace$ as:
\begin{equation}
\label{eq:10}       %{eq:14}
\mathbf n=n_i \mathbf{Y}_i+\cos{\theta} \hat{\mathbf N}\,,
\end{equation}
where the director components $n_i,~i=1,2$ and the tilt angle $\theta$ are functions of the surface coordinates $u_1,u_2$ (see Fig.~1).

\begin{figure}
\label{fig:1}
\centering
\includegraphics[width=0.9\columnwidth]{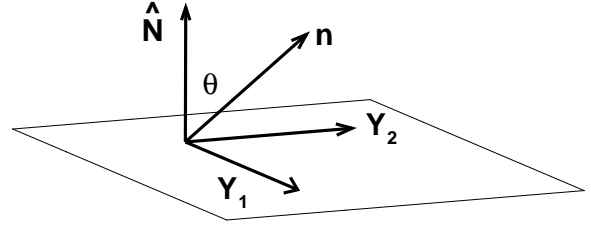}
\caption{The local membrane basis showing the tangent vectors $\mathbf{Y}_1$ and $\mathbf{Y}_2$ and the normal vector $\hat{\mathbf N}$ perpendicular to them. The director $\mathbf n$ is tilted by an angle $\theta$ with respect to $\hat{\mathbf N}$.}
\end{figure}

The unit length of the director field yields the constraint:
\begin{equation}
\label{eq:11}         %{eq:15}
 g_{ij}n_i n_j-\sin^2\theta=0\,.
\end{equation}
\smallskip
The three--dimensional gradient operator is given by
\begin{equation}
\label{eq:12}        %{eq:16}
\nabla=g^{ij} \mathbf{Y}_i \partial_j+\hat{\mathbf N}\partial_N\,,
%\quad\text{with}\quad\nabla'\equiv g^{ij} \vec{Y}_i \partial_j\,,
\end{equation}
where $g^{ij} \mathbf{Y}_i \partial_j$ describes the components of the gradient operator on the surface. To evaluate Eq.~\eqref{eq:8}, we need to evaluate the curl and the divergence of $\mathbf n$. The curl of $\mathbf n$ is calculated as
\begin{equation}
\label{eq:13}     %{eq:17}
\begin{split}
\nabla\times\mathbf n=\frac{\epsilon_{3ji}}{\sqrt{g}}&\left[\left(g_{ik}\partial_j n_k+g_{il} n_k \Gamma^l_{jk}\right)\hat{\mathbf N}\right. \\&\left.-\left(2 n_k L_{jk}+\partial_j \cos{\theta}\right)\mathbf{Y}_i\right]\,,
\end{split}
\end{equation}
where $\epsilon_{ijk}$ is the antisymmetric Levi--Civita tensor.
Using Eqs.~\eqref{eq:10} and~\eqref{eq:13} we obtain the twist of the director,
\begin{equation}
\label{eq:14}     %{eq:18}
\begin{split}
\mathbf n\cdot(\nabla\times\mathbf n)=\frac{\epsilon_{3ji}}{\sqrt{g}}&\left[\cos{\theta}\left(g_{ik}\partial_j n_k+g_{il} n_k \Gamma^l_{jk}\right)\right.\\&\left.-\left(2 n_k n_l g_{il} L_{jk}+g_{il} n_l\partial_j\cos{\theta}\right)\right]\,.
\end{split}
\end{equation}
The divergence of the director is given by
\begin{equation}
\label{eq:15}      %{eq:19}
\begin{split}
\nabla\cdot\mathbf n&=\partial_j n_j+n_l \Gamma^j_{jl}-\cos{\theta} L_{ji} g^{ij}\,,
\\&=\partial_j n_j+n_l \Gamma^j_{jl}\,,
\end{split}
\end{equation}
where we have used the fact that $H=\frac{1}{2}g^{ij}L_{ij}=0$ for a twisted ribbon
Using Eqs.~\eqref{eq:6},~\eqref{eq:13},~\eqref{eq:14},~\eqref{eq:15}, the free energy Eq.~\eqref{eq:7} is given by,
\begin{equation}
\label{eq:16}      %{eq:20}
\begin{split}
F=&\frac{K}{2}\int \left\lbrace(\partial_j n_j+n_l \Gamma^j_{jl})^2\right.
\\&\left.-2q \frac{\epsilon_{3ji}}{\sqrt{g}}\left[\cos{\theta}\left(g_{ik}\partial_j n_k+g_{il} n_k \Gamma^l_{jk}\right)\right.\right.\\&\left.\left.-\left(2 n_k n_l g_{il} L_{jk}+g_{il} n_l\partial_j\cos{\theta}\right)\right]\right.
\\&+\left.\left(\frac{\epsilon_{3ji}}{\sqrt{g}}\left[\left(g_{ik}\partial_j n_k+g_{il} n_k \Gamma^l_{jk}\right)\hat{\mathbf N}\right.\right.\right.\\&\left.\left.\left.-\left(2 n_k L_{jk}+\partial_j \cos{\theta}\right)\mathbf{Y}_i\right]\right)^2+q^2\right\rbrace\sqrt{g} du_1 du_2
\\&+\frac{C}{2}\int \sin^2\theta \sqrt{g} du_1 du_2+\bar{k} \int\frac{L}{g}\sqrt{g} du_1 du_2\\&+\gamma\oint dl\,.
\end{split}
\end{equation}
As noted in Ref.~\cite{ZhongCan2}, the term proportional to $q$ in the equation above is identical in form (up to a line integral that can be absorbed into the edge energy term proportional to $\gamma$) to the Helfrich--Prost term  \cite{HP} first introduced to model chirality in lipid membranes.

\subsection{\label{ribbon}S\lowercase{m}-$A^\ast$ twisted ribbon with Gaussian curvature term}

The position vector of a twisted ribbon (Fig.~2) of radius $R$ is given by
\begin{equation}
\label{eq:17}   % {eq:21}
\mathbf{Y}=\left\lbrace \rho\cos{\phi}, \rho\sin{\phi}, b\phi\right\rbrace\,, 
\end{equation}
where $|\rho|\leq R$ and $0\leq\phi\leq 2n\pi$, and $n$ is the winding number. The pitch of the twisted ribbon is given by $2\pi|b|$, and the sign of $b$ determines the handedness of the ribbon. Here we focus on a right-handed ribbon with $b>0$ without any loss of generality.

Defining $u_1\equiv\rho$ and $u_2\equiv\phi$, the tangent vectors and the normal vector of the surface are given by
\begin{equation}
\label{eq:18}      %{eq:22}
\begin{split}
&\mathbf{Y}_1=\partial_{\rho}\mathbf{Y}=\left\lbrace\cos{\phi}, \sin{\phi}, 0\right\rbrace\,,
\\&\mathbf{Y}_2=\partial_{\phi}\mathbf{Y}=\left\lbrace-\rho\sin{\phi}, \rho\cos{\phi}, b\right\rbrace\,,
\\&\hat{\mathbf N}=\frac{1}{\sqrt{g}} \left\lbrace b\sin{\phi}, -b\cos{\phi}, \rho \right\rbrace\,.
\end{split}
\end{equation}
Using Eq.~\eqref{eq:18} in Eq.~\eqref{eq:4} we find:
\begin{equation}
\label{eq:19}     %{eq:23}
\begin{split}
&g_{11}=g^{11}=1\,,\quad g_{22}=(g^{22})^{-1}=\rho^2+b^2\,,\quad g_{12}=g_{21}=0\,,
\\&L_{12}=L_{21}=\frac{-b}{\sqrt{\rho^2+b^2}}\,,\quad L_{11}=L_{22}=0\,,
\\&g=\rho^2+b^2\,, \quad L=-\frac{b^2}{\rho^2+b^2}\,,
\\&\Gamma_{12}^{2}=\frac{\rho}{\rho^2+b^2}\,,\quad \Gamma^1_{22}=-\rho\,;\quad 0\text{ otherwise}\,,
\\&H=0\,,\quad K_G=-\frac{b^2}{(\rho^2+b^2)^2}\,.
\end{split}
\end{equation}

As we will ultimately carry out a mean--field analysis of the free energy, we assume that the director field exhibits azimuthal symmetry. Furthermore we assume that the director is tilted parallel to the edge of the ribbon. The director field $\mathbf n$ is then given by
\begin{equation}
\label{eq:20}        
\mathbf n=\left\lbrace 0, \frac{\sin{\theta(\rho)}}{\sqrt{g}}, \cos{\theta(\rho)}\right\rbrace\,,
\end{equation}
where we have used the unit length constraint, Eq.~\eqref{eq:11}. Note that the director field of Eq.~\eqref{eq:20} has zero splay; this can be verified explicitly using Eq.~\eqref{eq:15}.

Within the single Frank constant approximation we have one penetration length given by $\lambda_t\equiv(\frac{K}{C})^{1/2}$. 
The thickness of the membrane is included in the elastic moduli $K$ and $C$. Inserting Eqs.~\eqref{eq:18},~\eqref{eq:19}, and~\eqref{eq:20} into Eq.~\eqref{eq:16} we find the following expression of the free energy:
\begin{equation}
\begin{split}
\label{eq:21}     %{eq:25}
F&=\int\frac{1}{2}\left[K\left(\frac{d \theta}{d \rho}-q\right)^2+K\left(\frac{\rho}{g} \sin{2 \theta}+\frac{4b}{g}\sin^2{\theta}\right)\right.\\&\left.\times\left(\frac{d \theta}{d \rho}-q\right)+K(g+3b^2)\left(\frac{1}{g}\sin{\theta}\right)^2\right.\\&\left.+C\sin^2{\theta}-2 \bar{k}\frac{b^2}{g^2}\right]\sqrt{g} d\rho d\phi+\gamma\int \sqrt{R^2+b^2}d\phi\,.
\end{split}
\end{equation}

The Euler--Lagrange (EL) equation for $\theta$ is given by:
\begin{equation}
\label{eq:23}    %{eq:27}
\begin{split}
2 g &\frac{d^2 \theta(\rho)}{d \rho^2}+2 \rho \frac{d \theta(\rho)}{d \rho}-\left(\frac{2 b^2}{g}+\frac{g}{\lambda_t^2}-4b q+1\right)\sin{2 \theta(\rho)}\\&-4\left(\rho q+\frac{ \rho b}{g}\right)\sin^2{\theta(\rho)}=0\,.
\end{split}
\end{equation}
The boundary conditions are given by:
\begin{eqnarray}
  %{eq:28}
&&\label{BC1} \theta(0)=0\\ 
&&2(R^2+b^2)\left(\frac{d \theta(\rho)}{d \rho}\bigg|_{\rho=R}-q\right)+\rho \sin{2 \theta(R)}\nonumber\\&&+4 b \sin^2{\theta(R)}=0.\label{BC2}  
\end{eqnarray}
Eq.~\eqref{BC1} arises from  the relation $\theta(\rho)=-\theta(-\rho)$ on a twisted ribbon, and is also observed experimentally, while Eq.~\eqref{BC2} is the free boundary condition, $\frac{\partial f_n}{\partial \theta'}=0$, where %$f_n$ is the dimensionless free energy density $f_n\equiv\frac{\sqrt{g}f_n}{K}$ and 
$\theta'\equiv\frac{\partial f}{\partial \theta}$. Physically, this boundary condition ensures that the director torque vanishes at the edge of the ribbon. The director field given by the solution of Eq.~\eqref{eq:23}, subject to the above boundary conditions, is shown in Fig.~2.

\begin{figure}
\label{fig:2}
\centering
\includegraphics[width=0.9\columnwidth]{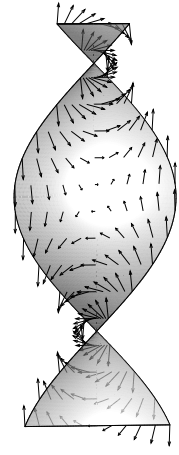}
\caption{A twisted ribbon with radius $R$ showing the director field determined by the solution of the Euler--Lagrange equation, Eq.~\eqref{eq:23}.  The figure shows a length of the ribbon equal to 0.75 times the pitch. }
\end{figure}

The EL equation and boundary conditions derived in Ref.~\cite{Barry} for a semi--infinite smectic layer (Fig.~3) with a straight boundary can be recovered from the above results for the twisted ribbon by taking the limit $b\rightarrow\infty$. On the other hand, the limit $b\rightarrow0$ with $\bar{k}=0$ and $R$ constant results in a finite disk of radius $R$. This limit leads to the EL equation and boundary conditions for a finite disk given in Ref.~\cite{Pelcovits}.

\begin{figure}
\label{fig:3}
\centering
\includegraphics[width=0.7\columnwidth]{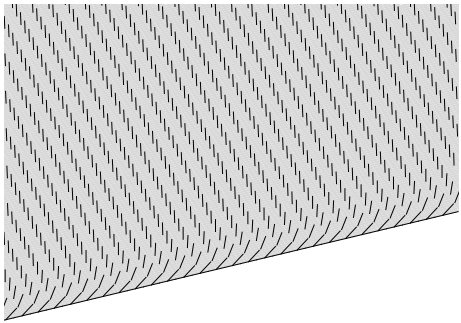}
\caption{Director field near the edge of a semi--infinite smectic layer, the limit of a twisted ribbon with  $b\rightarrow\infty$.  }
\end{figure}

\subsection{\label{ss}S\lowercase{m}-$A^\ast$ twisted ribbon with saddle--splay term}

We now consider a modification of the above free energy of a ribbon replacing the Gaussian curvature term in the Helfrich energy Eq.~\eqref{eq:2} by the Frank saddle--splay energy density, i.e, we replace the surface normal $\hat{\mathbf N}$ by the director field $\mathbf n$ in Eq.~\eqref{KG} and obtain the saddle--splay free energy density:

\begin{equation}
\label{eq:25}    %{eq:29}
f_{24}=\frac{K_{24}}{2} \nabla\cdot\left(\mathbf n(\nabla\cdot\mathbf n)-(\mathbf n\cdot\nabla)\mathbf n\right)\,,
\end{equation}
where $K_{24}$ is the saddle--splay modulus. Two major simplifications help us evaluate Eq.~\eqref{eq:25}: (1) the director given in Eq.~\eqref{eq:20} has no radial component; (2) due to our choice of the coordinate system, metric tensor $g_{ij}$ of the surface given in Eq.~\eqref{eq:19} is diagonal. Hence, Eq.~\eqref{eq:12} simplifies to
\begin{equation}
\label{eq:26}  %{eq:30} 
\nabla=g^{11}\mathbf{Y}_1 \partial_1+g^{22}\mathbf{Y}_2 \partial_2+\hat{\mathbf N}\partial_N\,.
\end{equation}
Calculating the first term on the right--hand side of Eq.~\eqref{eq:25} apart from the factor $\frac{K_{24}}{2}$, we obtain
\begin{equation}
\label{eq:27}           %{eq:32}
 \nabla\cdot\left(\mathbf n(\nabla\cdot\mathbf n)\right)=-4H^2\cos^2{\theta}+2 K_G \cos^2{\theta}\,,
\end{equation}
while the second term of Eq.~\eqref{eq:25} yields
\begin{equation}
\label{eq:28}       %{eq:31}
\begin{split}
 \nabla\cdot\left((\mathbf n\cdot\nabla)\mathbf n\right)=&\partial_1 \left((n_2)^2\Gamma_{22}^1\right)-2\partial_1\left(\cos{\theta} n_2 L_{21}\right)\\&+(n_2)^2 \Gamma_{22}^1\partial_2{Y}_1-2\cos{\theta}n_2 L_{21} \Gamma_{21}^2\\&-(n_2 L_{21})^2\,,
\end{split}
\end{equation}
with the director component $n_2$ given by Eq.~\eqref{eq:20}.
%Defining $K_{24}\equiv\frac{\tilde{K}_{24}}{\tilde{K}}$ and $f_{24}\equiv\frac{f_{24}}{\tilde{K}}$, 
Finally, the saddle--splay free energy density, Eq.~\eqref{eq:25} is given by:
\begin{equation}
\label{eq:29}     %{eq:33}
\begin{split}
f_{24}=&\frac{K_{24} \sqrt{g}}{2 g^2}
   \left(-2 b^2 \cos ^2\theta+2\sin\theta
   \left(b^2\sin\theta+\rho g\cos\theta \frac{\partial \theta}{\partial \rho}\right)\right.\\&\left.-b \left(2 g \cos{2\theta}\frac{\partial \theta}{\partial \rho}-\rho\sin{2 \theta}\right)\right)\,.
\end{split}
\end{equation}
In the limit of zero tilt angle (i.e., $\hat{\mathbf n}=\hat{\mathbf N}$), Eq.~\eqref{eq:29} reduces to the term proportional to $\bar{k}$ in Eq.~\eqref{eq:21} if we replace $K_{24}$ by $\bar{k}$. The EL equation including the saddle--splay term in place of the Gaussian curvature is given by:
\begin{equation}
\label{eq:30}      %{eq:34} 
\begin{split}
&-2 g\left(\rho \frac{\partial \theta}{\partial \rho}+g \frac{\partial^2 \theta}{\partial \rho^2}\right)+4 \rho \left(q b^2+b+q\rho^2\right) \sin ^2\theta\\&+\left[\frac{\rho^4}{\lambda_t^2}+\left(2 b (\frac{b}{\lambda_t^2}-2 q)+1\right) \rho^2\right.\\& \left.+b^2 \left(\frac{b^2}{\lambda_t^2}-4qb+3 \frac{K_{24}}{K}+3\right)\right] \sin{2 \theta}=0\,,
\end{split}
\end{equation}
with the boundary condition at the edge $\rho=R$:
\begin{equation}
\label{eq:31}       %{eq:35}
\begin{split}
&-2 \left(q R^2+b (b q-1)\right)-2 b (\frac{K_{24}}{K}+1)
   \cos{2 \theta}\\&+(\frac{K_{24}}{K}+1) R \sin{2 \theta}+2(R^2+b^2) \frac{\partial \theta}{\partial \rho}\bigg|_{\rho=R}=0\,.
\end{split}
\end{equation}

\section{Results}
\label{results}

\begin{figure}
\label{fig:4}
\centering
\includegraphics[width=1\columnwidth]{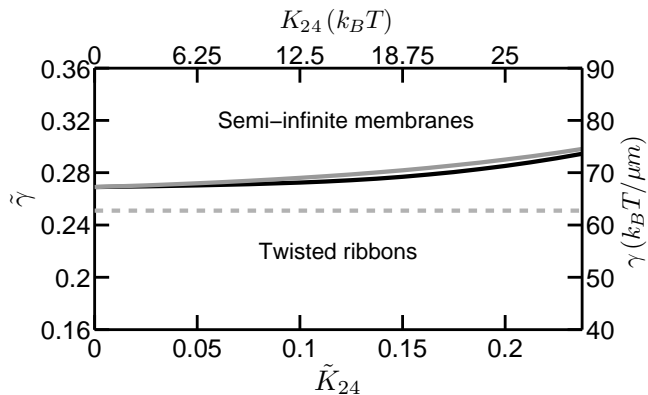}
\caption{Phase diagram for semi--infinite membranes and twisted ribbons with $q\lambda_t=0.71$. The results are shown both in dimensionless units (left and bottom axes), $\tilde{K}_{24}\equiv\frac{K_{24}}{K}$ and $\tilde{\gamma}\equiv\frac{\gamma \lambda_t}{K}$, and physical units (top and right axes) obtained using the value of $\lambda_t$ measured in \cite{Pelcovits} and the value of $K$ measured in \cite{Dogic1}. The solid curves denote first--order phase transition lines for the Gaussian curvature (black) and saddle--splay (gray) models. The saddle--splay modulus $K_{24}$ and the Gaussian modulus $\bar{k}$ are taken to be equal.  The dashed line is \textit{not} a phase boundary; rather it denotes the second--order phase transition between finite--sized disks and semi--infinite membranes obtained in a theoretical model \cite{Pelcovits} which did not consider twisted ribbons. The twisted ribbon is of lower energy than the finite--sized disks.}
\end{figure}

\begin{figure*}
\label{fig:5}
%\centering
%\includegraphics[width=0.85\textwidth]{angle_paper.eps}
\includegraphics[scale=0.6]{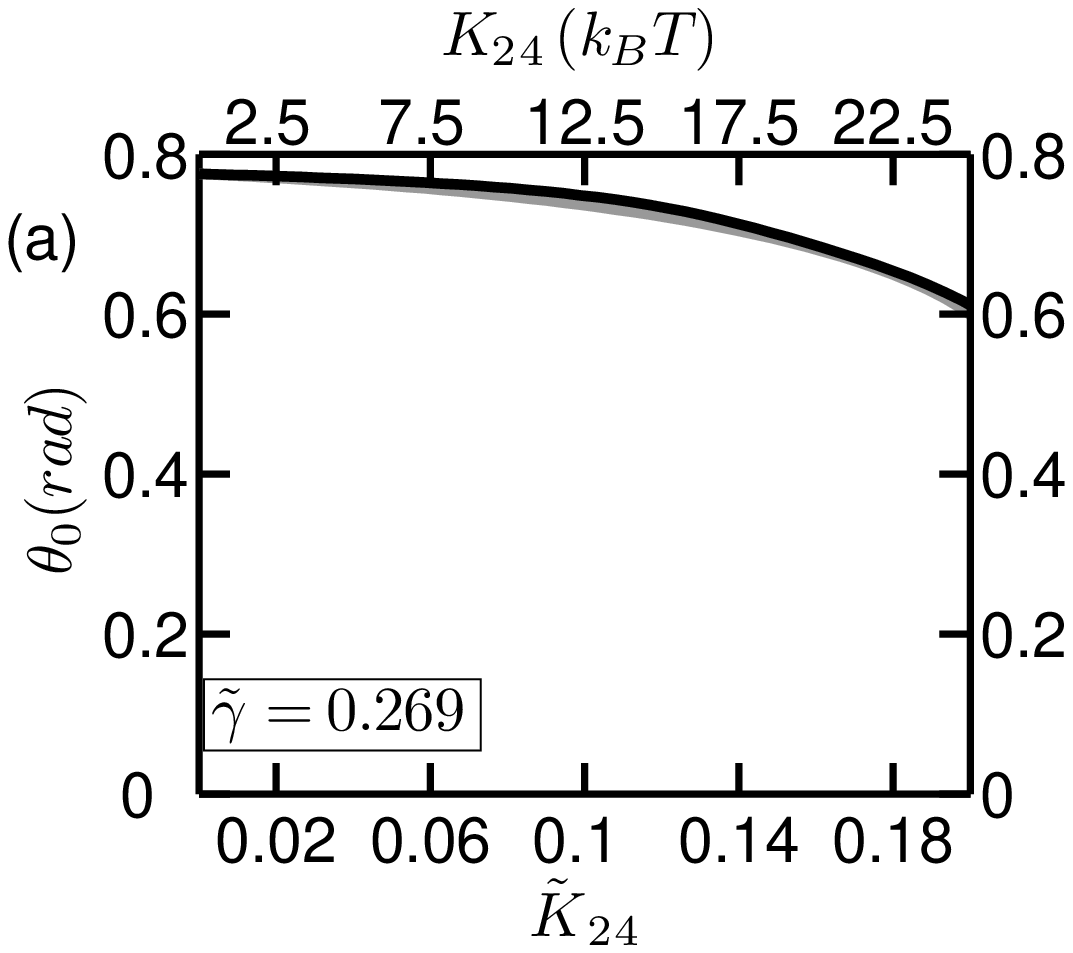}
\includegraphics[scale=0.6]{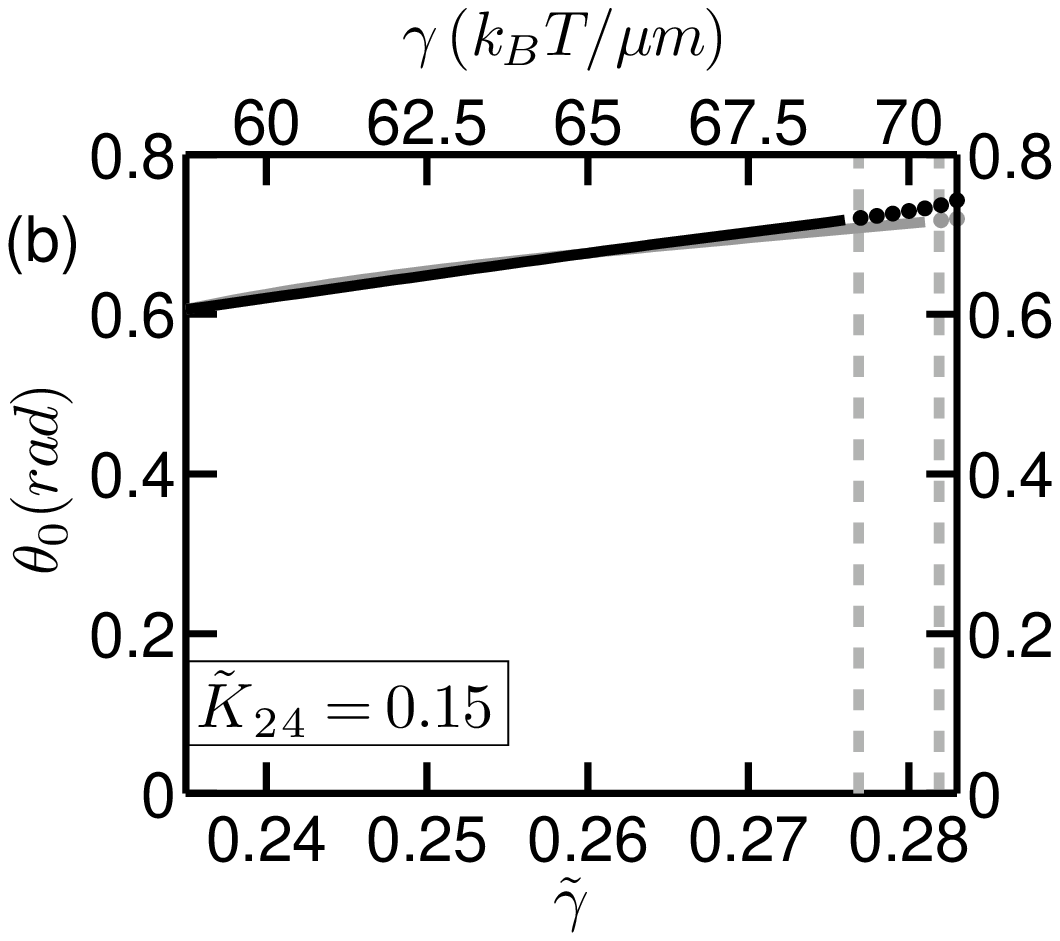}
\includegraphics[scale=0.6]{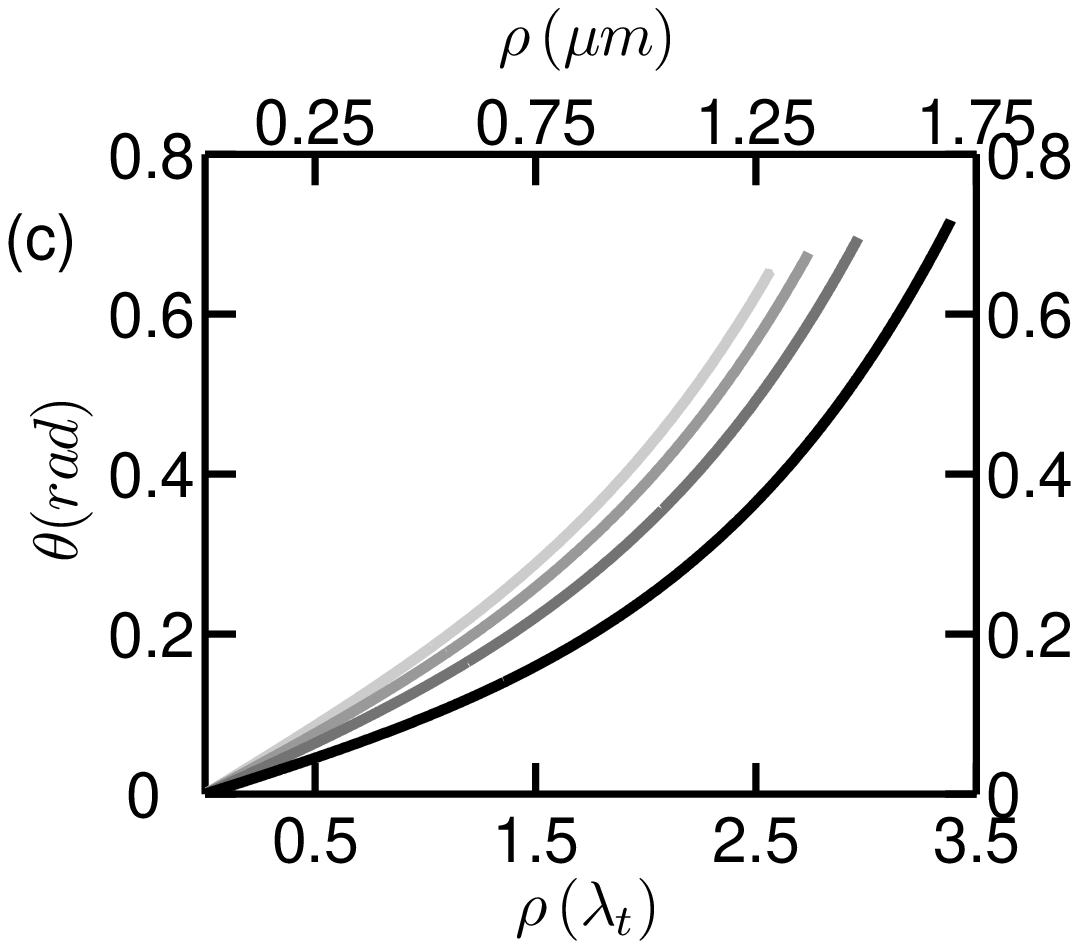}
\caption{(a) The value of the tilt angle $\theta_0\equiv\theta(R)$ at the edge of the ribbon as a function of $\tilde{K}_{24}$ for  $\tilde{\gamma}=0.269$, corresponding to a point slightly below the first--order phase boundary shown in Fig.~4. The black and gray curves correspond to the Gaussian curvature and saddle--splay models respectively. The values of $\theta_0$ shown are all less than the value found for the semi--infinite membrane \cite{Barry}; (b) $\theta_0$ as a function of $\tilde{\gamma}$ for $\tilde{K}_{24}=0.15$, as the phase boundary  is approached from below. The vertical dashed lines denote from left to right the first--order phase transition from twisted ribbons to semi--infinite membranes for the Gaussian curvature and saddle splay models. To the right of these lines the ribbon is metastable; this is indicated by the dashed nature of the $\theta_0$ curves. (c) The tilt angle $\theta(\rho)$ for the saddle--splay model with  $\tilde{K}_{24}=0.15$ as a function of $\rho$ in units of $\lambda_t$ (lower axis) and physical units (top axis) corresponding from top to bottom to the parameters:  $\tilde{\gamma}=0.25$, ($R=2.566\lambda_t$, $b=2.877\lambda_t$);  $\tilde{\gamma}=0.26$, ($R=2.739\lambda_t$, $b=3.381\lambda_t$,);  $\tilde{\gamma}=0.27$, ($R=2.965\lambda_t$, $b=3.899\lambda_t$); and $\tilde{\gamma}=0.28187$, ($R=3.385\lambda_t$, $b=4.707\lambda_t$,). The lowest curve ($\tilde{\gamma}=0.28187$) corresponds to the phase boundary.}
\end{figure*}

\begin{figure*}
\label{fig:6}
\includegraphics[scale=0.65]{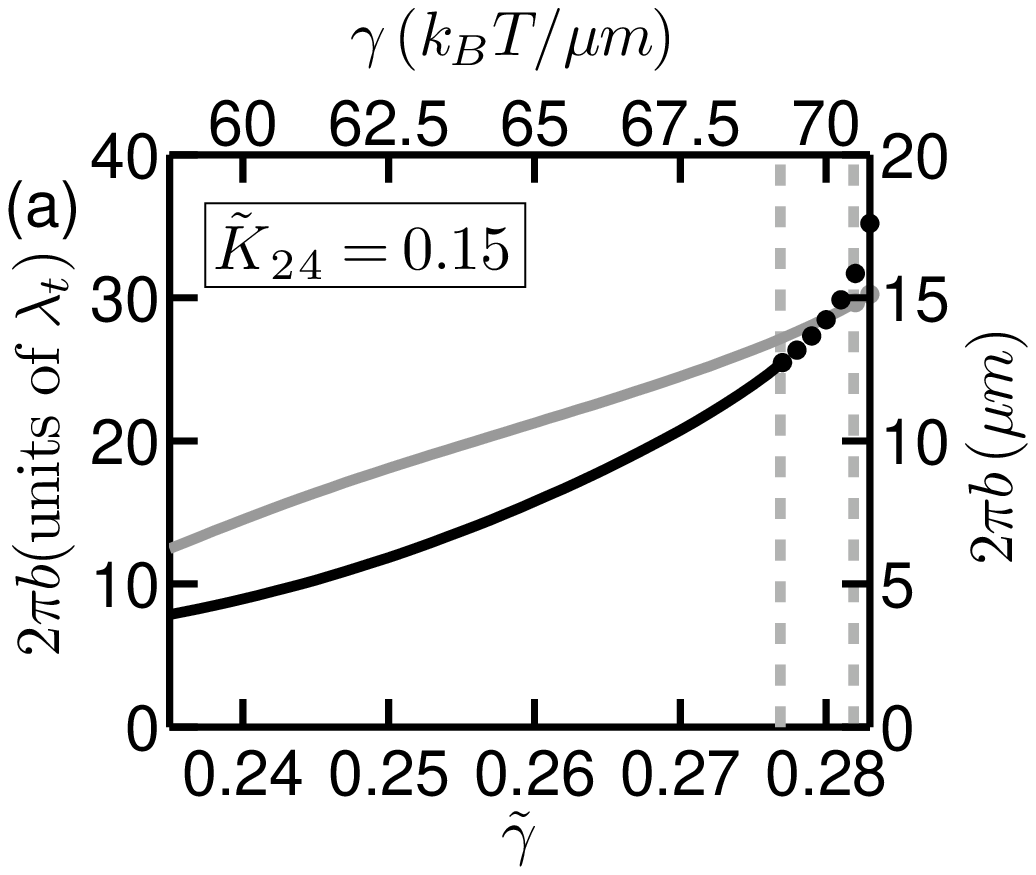}
\includegraphics[scale=0.65]{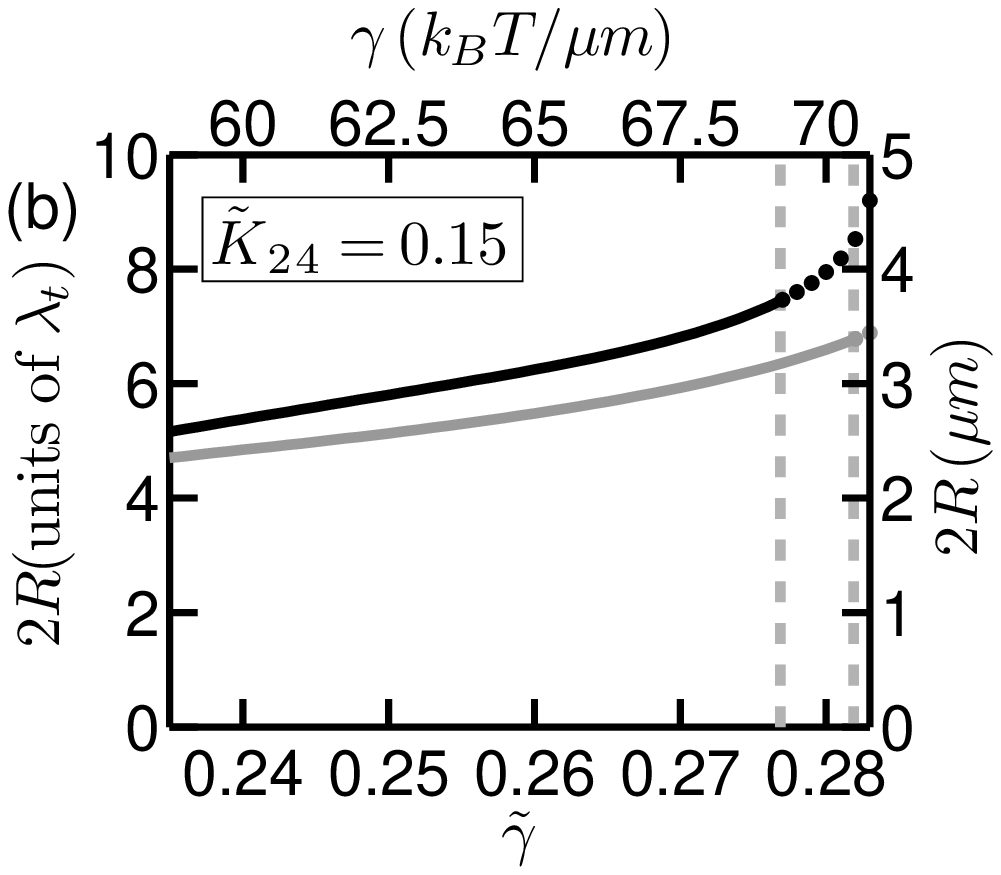}
\includegraphics[scale=0.6]{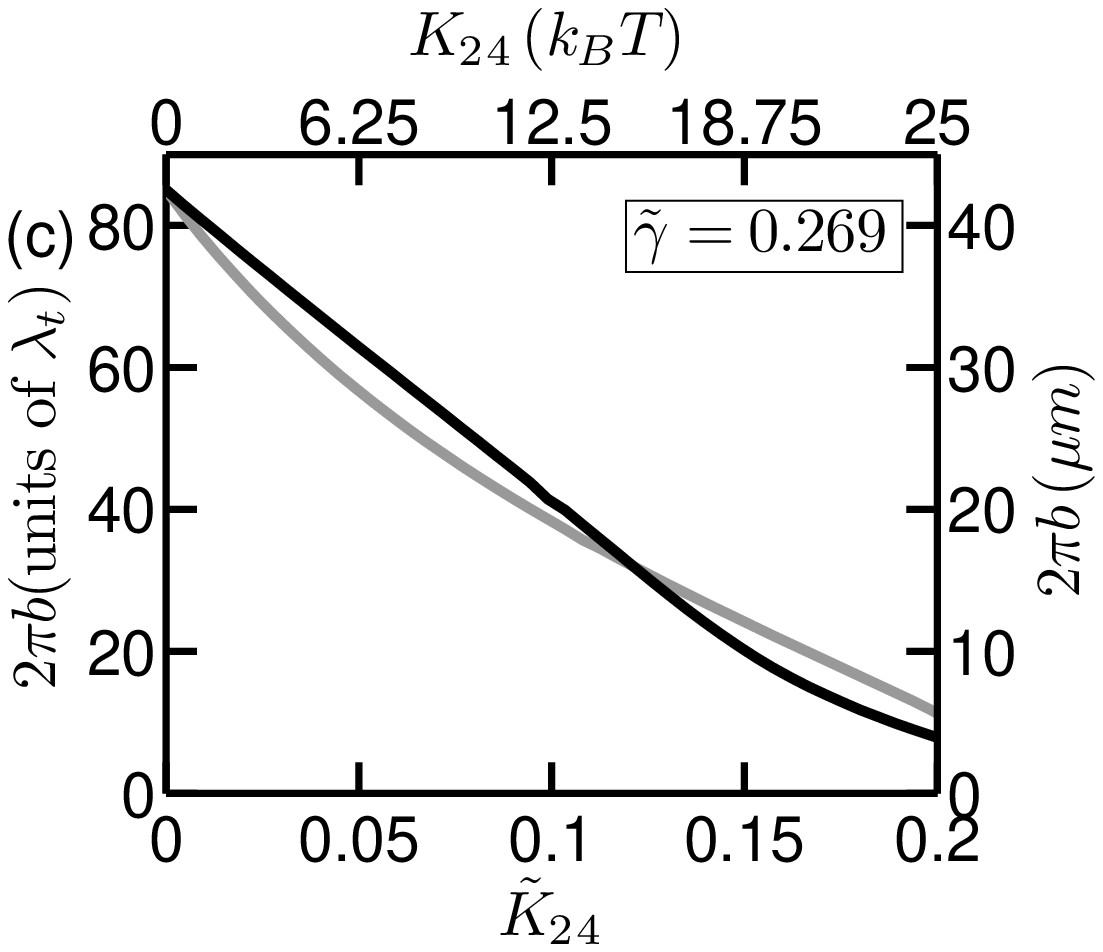}
\includegraphics[scale=0.6]{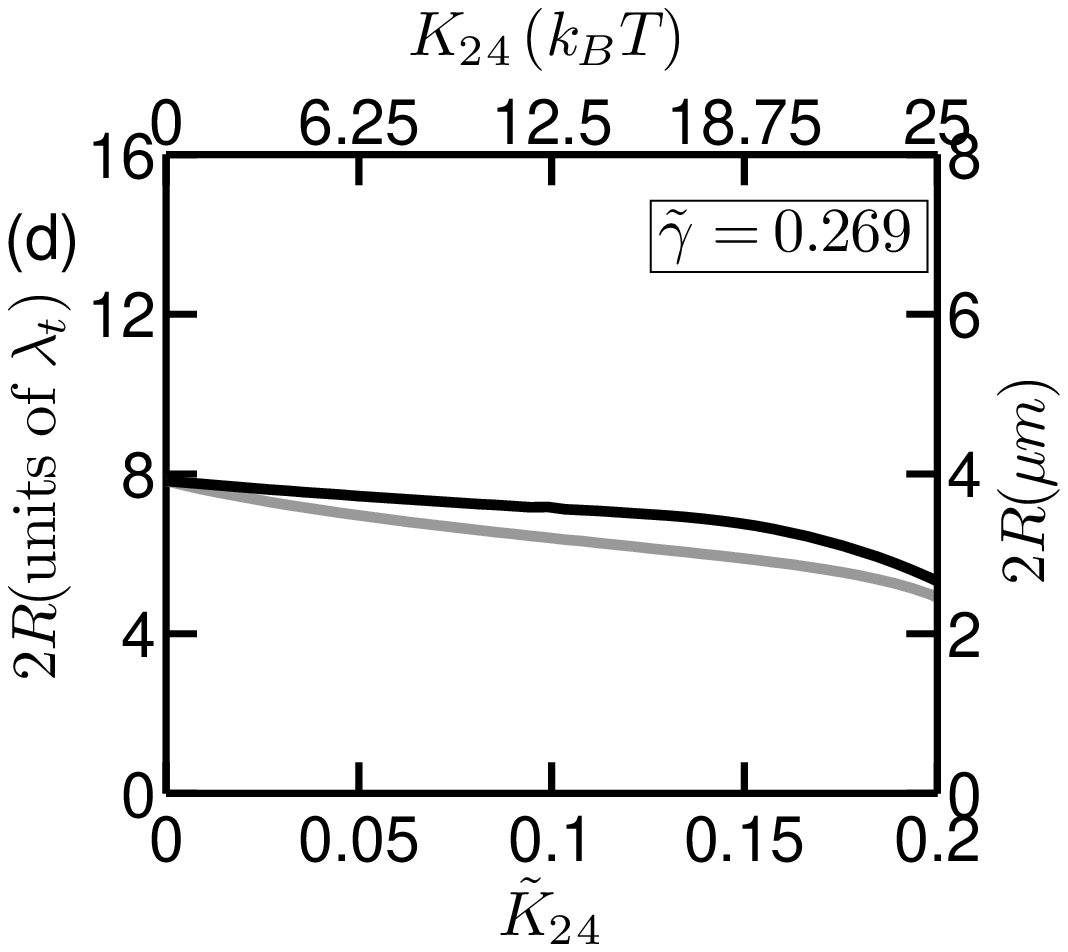}
\includegraphics[scale=0.65]{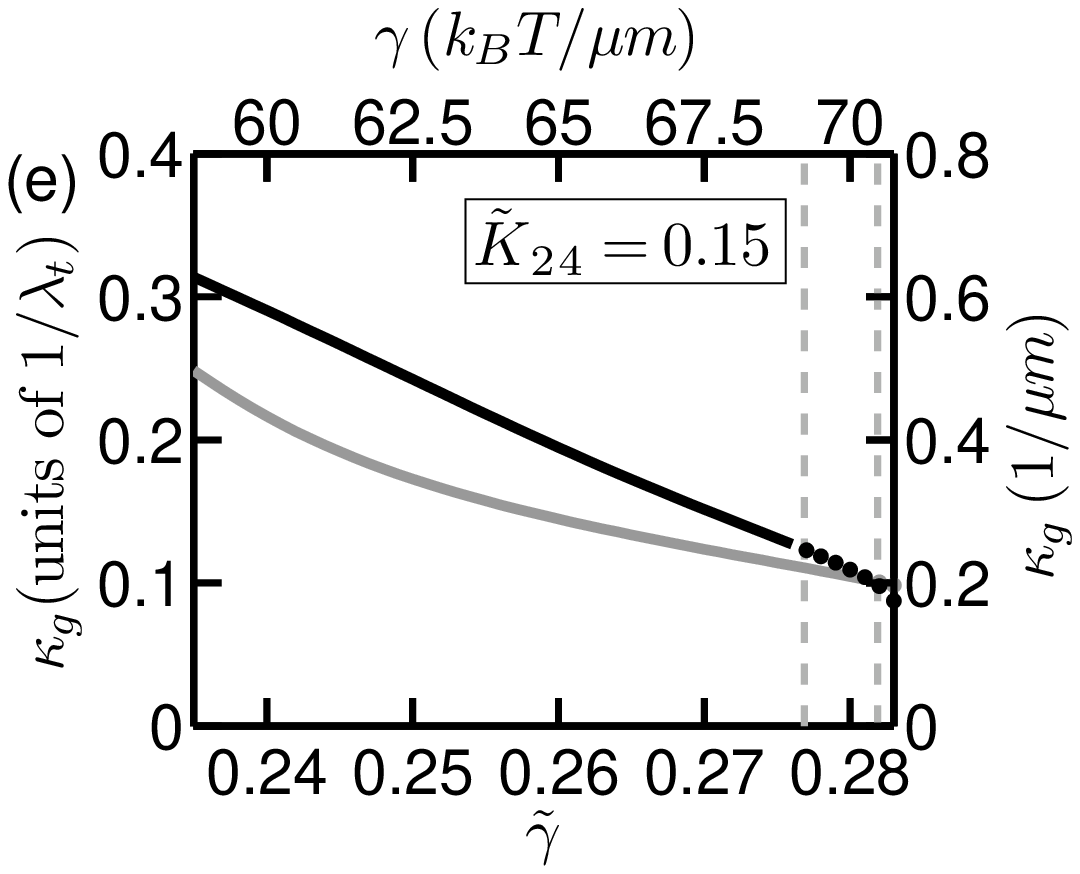}
\caption{The (a) pitch $2\pi b$ and (b) width $2R$ of the ribbon in units of $\lambda_t$ (left vertical axis) and $\mu m$ (right vertical axis) as functions of $\tilde{\gamma}$ at $\tilde{K}_{24}=0.15$. Black and gray curves correspond to the Gaussian curvature and saddle--splay models respectively.  The gray vertical dashed lines denote from left to right the first--order phase transition from ribbons to semi--infinite membranes for the Gaussian curvature and saddle--splay models, respectively. To the right of these lines the ribbon is metastable; this is indicated by the dashed nature of the curves. In (c) and (d) respectively the pitch and width are shown as functions of $\tilde{K}_{24}$ close to the phase boundary at $\tilde{\gamma}=0.269$. Again, black and gray curves correspond to the Gaussian curvature and saddle--splay models respectively.  The geodesic curvature, Eq.~\eqref{kappag}, as a function of $\tilde{\gamma}$ is shown in (e). } %The phase transition of the saddle--splay model lies close to the second dashed line, and is not shown.}
\end{figure*}

We now explore the results of solving the EL equations and their respective boundary conditions for the two models of the twisted ribbon: with the Gaussian curvature term (Eqs.~\eqref{eq:23}--\eqref{BC2}) and the saddle--splay energy (Eqs.~\eqref{eq:30}--\eqref{eq:31}). We ignore any change in the membrane thickness due to the tilt of the director and assume a common value of the Gaussian modulus $\bar{k}$ and the saddle--splay modulus $K_{24}$ in comparing the two models. For a given value of these parameters and $\gamma$, the edge energy, we determine, using a downhill--simplex algorithm, the values of the radius $R$ and pitch $b$ which minimize the free energy of the ribbon. We then compare the free energy per unit area of the twisted ribbon with the corresponding energy density of a semi--infinite Sm-$A^\ast$ membrane \cite{Barry}, where as in the case of the ribbon, the director is tilted parallel to the edge of the membrane (Fig.~3). We determine the phase boundary between the two structures using a bisection algorithm. The area $A_{r}$ of a twisted ribbon is given by
\begin{equation}
\label{eq:32}
\begin{split}
A_{r}&=\int \sqrt{g} d\rho d\phi
=2\pi \int^R_{-R} \sqrt{\rho^2+b^2} d\rho\\&=2\pi b^2\left[\frac{R}{b} \sqrt{1+\left(\frac{R}{b}\right)^2}+\sinh^{-1}\left(\frac{R}{b}\right)\right]\,.
\end{split}
\end{equation}
%Referring to Eq.~\ref{eq:21}, 
The energy per unit area of the semi--infinite membrane is given simply by $q^2/2$ as director tilting occurs only in a small region near the edge, and thus in Eq.~\eqref{eq:1} the contributions of twist and bend deformations to the mean free energy density are negligible.

We have considered both positive and negative values of the moduli $\bar{k}$ and $K_{24}$. We find that negative values of these quantities lead to stable ribbons with ratios of pitch to radius of order 35 or greater, significantly larger than what is measured experimentally \cite{Gibaud}. On the other hand, positive values of the moduli lead to stable ribbons with ratios of pitch to radius of order 5, in good agreement with experiment. Positive values of these moduli are not typically measured in lipid monolayers or bilayers \cite{Marsh}; however, there is no reason to exclude this possibility on physical grounds and the $fd$ system may be very different from systems composed of amphiphilic molecules. Positive moduli lead to saddle--splay and Gaussian curvature energies which are negative for a twisted ribbon. Thus, to stabilize the free energy we have added a higher--order curvature term proportional to $K_G^2$ \cite{Shemesh, Mitov}. The coupling of this term to the free energy can be chosen as small as $0.01$ in dimensionless units to guarantee stability for the range of $K_{24}$ and $\bar{k}$ values we have studied.  

Our main results are displayed in Figs.~4--6, where we present the phase diagram for semi--infinite membranes and twisted ribbons, and values of the tilt angle, pitch and width of the ribbon as functions of dimensionless variables $\tilde{K}_{24}\equiv\frac{K_{24}}{K}$ and $\tilde{\gamma}\equiv\frac{\gamma \lambda_t}{K}$, as well as physical units obtained by using the value of $K$ measured in \cite{Dogic1} and  $\lambda_t$ measured in \cite{Pelcovits}.  Our results were obtained for $q\lambda_t=0.71$, corresponding to the values of these parameters measured experimentally in Ref.~\cite{Pelcovits}. We find a first order phase transition between ribbons and semi--infinite membranes (Fig.~4) as the edge energy $\gamma$ (proportional to the depletant concentration) is varied. Furthermore, we see that even though there is a finite tilt of the molecules throughout the layer (see Fig.~5c) which should in principle distinguish between the saddle--splay and Gaussian curvature models, there is no significant difference between the calculated phase boundaries for the two models. The dashed line in Fig.~4 is \textit{not} a phase boundary; rather it is the second--order phase transition line between semi--infinite membranes and finite--sized disks calculated in a model \cite{Pelcovits} which excluded the possibility of twisted ribbons. The present calculation demonstrates that ribbons are of lower free energy than the finite--sized disks and the first--order phase transition from semi--infinite membranes to ribbons preempts the transition to finite--sized disks as $\gamma$ is reduced, in agreement with experimental observations \cite{Gibaud}.  An examination of the various contributions to the free energy indicates that it is the reduction in the chiral energy density $-qK\mathbf n\cdot\left(\nabla\times \mathbf n\right)$ that drives the transition to the ribbon state. If $q$, the measure of chirality, is set equal to zero, ribbons are never the lowest free energy structure.

The upward slope of the phase boundary in Fig.~4 indicates  that as $\tilde{K}_{24}$ increases the ribbon is energetically more favorable than the semi--infinite membrane. This tendency arises because increasing $\tilde{K}_{24}$ makes the saddle--splay energy (or equivalently the Gaussian curvature energy) more negative, thus making the ribbon more favorable. This effect outweighs two factors making ribbons less favorable as $\tilde{K}_{24}$ increases, namely,  the bend energy increases and there is less twist penetration at the edge. The increase in bend energy can be inferred from Fig.~6c where the ribbon pitch $b$ is plotted as a function of $\tilde{K}_{24}$ for fixed $\tilde{\gamma}$. Recalling (see the end of Sec.~\ref{ribbon}) that the limit $b\rightarrow\infty$ corresponds to a semi--infinite membrane (where the bend deformation vanishes) while the opposite limit $b\rightarrow 0$ corresponds to a disk of radius $R$ where the director field exhibits bend \cite{Pelcovits}, Fig.~6c indicates that the bend energy increases with increasing $\tilde{K}_{24}$. The decrease in twist penetrations as $\tilde{K}_{24}$ increases can be seen from  Fig.~5a where the tilt angle $\theta_0\equiv\theta(R)$ at the edge of the ribbon is plotted as a function of $\tilde{K}_{24}$. 

While Fig.~4 is in qualitative agreement with experiment, namely, that a first order transition is observed between semi--infinite membranes and twisted ribbons, the predicted line tension at the phase boundary, $\gamma\sim 10~k_B T/\mu m$ is an order of magnitude lower than the value measured in experiments \cite{Gibaud}. This discrepancy persists even when one accounts for the fact that the experiments measure an effective line tension, i.e., including the net reduction of the free energy near the edge due to twist penetration.

The tilt angle $\theta_0\equiv\theta(R)$ at the edge of the ribbon for both the Gaussian curvature and saddle--splay models is shown as a function of both $\tilde{K}_{24}$ and the edge energy $\gamma$ in Figs.~5a and 5b respectively. From Fig.~5a we see that as with the phase boundary shown in Fig.~4 there is negligible difference between the results for the two models. The vertical dashed lines in Fig.~5b correspond to the first--order phase transition boundaries shown in Fig.~4 for the saddle--splay and Gaussian curvature models. To the right of these lines the ribbon is metastable which we indicate by the dashed plotting of $\theta_0$. 
For semi--infinite membranes \cite{Barry} $\theta_0=-\arcsin(q \lambda_t)=0.79$ for $q \lambda_t=0.71$, independent of $\tilde{\gamma}$. This value exceeds $\theta_0$ for the twisted ribbon in both the saddle--splay and Gaussian curvature models for all values of $\tilde{\gamma}$ and $\tilde{K}_{24}$, in agreement with experiment \cite{Gibaud}.  In Fig.~5c we show examples of the tilt angle profile $\theta(\rho)$ for the saddle--splay model with $\tilde{K}_{24}=0.15$ for a range of $\tilde{\gamma}$ slightly below and on the first--order phase boundary of Fig.~4.

The pitch $2\pi b$ and width $2R$ of the ribbon are shown as functions of $\tilde{\gamma}$ and $\tilde{K}_{24}$ in Figs.~6a--d. In Figs.~6a and 6b we see that there is some difference between the Gaussian curvature and saddle--splay models for the dependence of the pitch on $\tilde{\gamma}$; we have no simple physical reason for this difference. Fig.~6e shows the geodesic curvature $\kappa_g$ \cite{Kamien} of the edge of the ribbon:
\begin{equation}
\label{kappag}
\kappa_g=\frac{R}{R^2+b^2}
\end{equation}
as a function of $\tilde{\gamma}$ in the neighborhood of the first--order transition. For both the Gaussian curvature and saddle--splay models the geodesic curvature decreases as the phase transition is approached from below, while Fig.~5b shows that the $\theta_0$ increases. This result is consistent with experimental retardance data which shows that lower curvature of the edge in general leads to a higher tilt angle at the edge.

\section{Conclusion}

In this paper, we have developed a theoretical model for the Sm-$A^\ast$ twisted ribbons observed in assemblies of $fd$ viruses condensed by depletion forces. The depletion interaction is modeled by an edge energy (line tension), assumed to be proportional to the depletant polymer in solution.  We have considered two variants of this model, one with the conventional Helfrich Gaussian curvature term, and a second with saddle--splay energy. Both models yield a first--order phase transition between ribbons and semi--infinite flat membranes as the edge energy is varied. The phase transition line and tilt angle profile are found to be nearly identical for the two models; the pitch of the ribbon, however, does show some differences. Our model yields good qualitative agreement with experimental observations, namely, (1) the existence of a first--order phase transition between ribbons and semi--infinite membranes, preempting a transition to finite--sized disks; (2) the dependence of the tilt angle at the edge of the ribbon on the curvature of the edge; (3) the decrease in the tilt angle at the edge of the membrane as the system undergoes the first--order transition from the semi--infinite membrane to the ribbon; (4) the order of magnitude of the pitch to width ratio of the ribbon. While the phase diagram is in qualitative agreement with experiment the predicted line tension on the phase boundary is an order of magnitude lower than that measured experimentally. Also, as we noted in Sec.~\ref{results}, obtaining the correct order of magnitude of the pitch to width ratio requires a positive value of the Gaussian curvature or saddle--splay modulus which favors negative Gaussian curvature. These moduli have not been directly measured in the $fd$ system nor have microscopic models been developed for nonamphiphilic molecules.

\begin{acknowledgments}
We thank E. Barry, Z. Dogic, T. Gibaud, and M. Zakhary for useful discussions. This work was supported by the NSF through MRSEC 0820492 and by Brandeis University.
\end{acknowledgments}


\medskip
\begin{thebibliography} {}
\bibitem{DeGennes1} P.~G. De Gennes and J. Prost, \textit{The Physics of Liquid Crystals}  (Oxford University Press, 1993).
\bibitem{DeGennes2} P.~G. De Gennes, Solid State Commun. {\bf 10,} 753 (1972).
\bibitem{Selinger}See, e.g.,  R.~D. Kamien and J.~V. Selinger, J. Phys: Cond. Mat. {\bf 13,} R1 (2001)
\bibitem{Barry} E. Barry, Z. Dogic, R.~B. Meyer, R.~A. Pelcovits, and R. Oldenburg,  J. Phys. Chem B {\bf 113,} 3910 (2009). 
\bibitem{Dogic3} E. Barry, D. Beller, and Z. Dogic, Soft Matter {\bf 5,} 2563 (2009).

\bibitem{Dogic1} Z. Dogic and S. Fraden, Langmuir {\bf 16,} 7820 (2000).

\bibitem{Dogic2} Z. Dogic and S. Fraden, Curr. Opin. Colloid Interface Sci. {\bf 11,} 47 (2006).
\bibitem{Pelcovits} R.~A. Pelcovits and R.~B. Meyer, Liq. Cryst. {\bf 36,}  1157 (2009).
\bibitem{Gibaud} T. Gibaud and Z. Dogic, private communication.
\bibitem{Helfrich1} W. Helfrich, Naturforsch. {\bf 28,} 693 (1973).
\bibitem{Helfrich2} W. Helfrich and H.~J. Deuling, J. Phys. (Paris) Colloq. {\bf 36,} C1327 (1975).
\bibitem{Seifert} Z.~C. Tu and U. Seifert, Phys. Rev. E {\bf 76,} 031603 (2007).
\bibitem{Kamien}See, e.g., R.~D. Kamien, Rev. Mod. Phys. {\bf 74,} 953 (2002).

\bibitem{Kleman} M. Kleman and O.~D. Lavrentovich, \textit{Soft Matter Physics: An Introduction}  (Springer Verlag, 2003).
%\bibitem{Lubensky} S.~R. Renn and T.~C. Lubensky, Phys. Rev. A {\bf 38,} 2132 (1988).
\bibitem{ZhongCan2} Ou-Yang Zhong-can and L. Ji-xing, Phys. Rev. Lett. {\bf 65,} 1679 (1990), Phys. Rev. A {\bf 43,} 6826 (1991).
\bibitem{ZhongCan1} Ou-Yang Zhong-can and W. Helfrich, Phys. Rev. A {\bf 39,} 5280 (1989). 
%\bibitem{Stoker} J.J. Stoker, \textit{Differential Geometry} (John Wiley\&Sons, Inc. 1969).
\bibitem{HP} W. Helfrich and J. Prost, Phys. Rev. A {\bf 38,} 3065 (1988).
\bibitem{Marsh} D. Marsh, Chem. and Phys. of Lipids, {\bf 144,} 146 (2006).
\bibitem{Shemesh} T. Shemesh, A. Luin, V. Malhotra, K.~N.~J. Boerger and M.~M. Kozlov, Biophys. J. {\bf 85,} 3813 (2003).
\bibitem{Mitov} M.~D. Mitov, Comptes Rendus Acad. Bulgares Sci {\bf 31,} 513 (1978).

\end{thebibliography}
\end{document}